\documentclass{epl}

\title{Frequency shifts of photoassociative spectra of ultracold metastable Helium atoms :
a new measurement of the s-wave scattering length}

\shorttitle{Frequency shifts of photoassociative spectra of $^4$He$^*$}

\author{J. Kim, S. Moal, M. Portier, J. Dugu\'{e}, M. Leduc \and C. Cohen-Tannoudji}

\institute{Ecole Normale Sup\'{e}rieure and Coll\`{e}ge de France, \\
Laboratoire Kastler Brossel, 24 rue Lhomond, 75231 Paris Cedex 05,
France.}




\begin{document}

\maketitle

\begin{abstract}
We observe light-induced frequency shifts in one-color photoassociative spectra of magnetically trapped $^4$He$^*$ atoms in the metastable $2^3S_1$ state. A pair of ultracold spin-polarized $2^3S_1$ helium atoms is excited into a molecular bound state in the purely long range $0_u^+$ potential connected to the $2^3S_1 - 2^3P_0$ asymptote. The shift arises from the optical coupling of the molecular excited bound state with the scattering states and the bound states of two colliding $2^3S_1$ atoms. We measure the frequency-shifts for several ro-vibrational levels in the $0^+_u$ potential and find a linear dependence on the photoassociation laser intensity. Comparison with a theoretical analysis provides a good indication for the s-wave scattering length $a$ of the quintet ($^5\Sigma_g^+$) potential, $a=7.2\pm 0.6$~nm,  which is significantly lower than most previous results obtained by non-spectroscopic methods.
\end{abstract}

\section{Introduction}
The present article describes a photoassociation (PA) experiment with ultracold helium atoms in the metastable ($\tau \sim$ 8000 s) $2^3S_1$ state. Since the development of laser cooling techniques which provide sub-millikelvin ultracold atomic samples, the photoassociation of such atoms has been of much interest and provides detailed information on the inter-atomic interaction and collisional properties~\cite{PA-Reviews}. The measurement of the PA spectra allows very precise measurement of the s-wave scattering length~\cite{PA-Li-a,PA-Na-a,PA-K-a,PA-Cs-a}, a crucial parameter for understanding the collisional properties of the ultracold atoms or molecules and the dynamic behavior of the condensate.

One of the interesting phenomena in PA is the light induced frequency shift of the PA spectra. As demonstrated in Ref.~\cite{PA-Na-BEC,Samuelis} for Na and Ref.~\cite{Gerton,Prodan} for Li, the light shift of the PA spectra is clearly visible at moderate laser intensity for the ultracold samples. The observed shifts are described by a theoretical calculation based on the theory developed by Bohn and Julienne~\cite{Bohn99} or Simoni \emph{et al.}~\cite{Simoni}. The laser light couples the excited molecular bound state with the continuum of scattering states and the ground molecular bound states, which results in the shift of the PA resonance curve when the laser frequency is swept over the resonance. The dependance of the shift on $a$ has two origins. First, in the limit of large and positive scattering length $a$, the energy of the least bound-state $E_{LBS}$ in the ground state potential vanishes, and varies as $-\hbar^2/2\mu a^2$. At ultracold temperature, the atoms collide with a near zero relative kinetic energy $E_{sc}$, which is therefore very close to $E_{LBS}$. When the laser is tuned in the vicinity of the free-bound transition, the excited state is therefore close to resonance with the least-bound state. As a consequence, a shift of the excited state is produced, which is dependent on $E_{LBS}$ and so on $a$. Second, the couplings of the excited state with the bound and scattering ground states involve Franck-Condon overlaps which depend critically on $a$.

\begin{figure}
\begin{center}
\resizebox{0.45\columnwidth}{!}{
\includegraphics{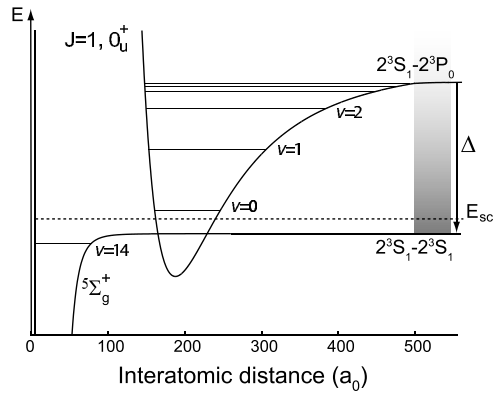}}
\end{center}
\caption{Interaction potential $^5\Sigma^+_g$ between spin-polarized metastable helium atoms and purely long-range $0^+_u$ potential in the light-dressed interaction picture. $\Delta$ is the detuning to the atomic $D_0$ line. Here, the colliding atoms at energy $E_{sc} \approx k_B T$ are driven close to resonance with the deepest bound state v=0 of $0^+_u$. This bound state is embedded in the continuum of scattering states, represented in grey, which leads to an energy shift. The least bound state v=14 of the $^5\Sigma^+_g$ potential, close to the resonance, also contributes to the shift.} \label{fig:potential}
\end{figure}

In this letter, we report the first observation of the light shift of the J=1 ro-vibrational states recently found by our group~\cite{PA-He,Kim} in the purely long-range $0_u^+$ interaction potential for $^4$He atoms. The v=0, 1, 2 levels (see Fig.~\ref{fig:potential}) can be populated from two ultracold colliding atoms excited by absorption of
laser light red-detuned from the $2^3S_1 - 2^3P_0$ transition. From the measurement of the light shifts for several vibrational levels, we deduce an estimation of the value of the
s-wave scattering length  $a$ of the spin polarized $^4$He$^*$ atoms in $^5\Sigma_g^+$ quintet potential. This is the first spectroscopic measurement of $a$, which provides a significantly lower value than most previous experimental results~\cite{Robert,He-BEC1,Tol, Seidelin}.


\section{Experimental Procedure}

The spin-polarized ultracold atomic $^4$He$^*$ samples are confined in a three-coil magnetic trap of Ioffe-Pritchard type and cooled down to $\mu$K range by radio-frequency-induced
evaporation. The bias field is typically 3 G and the density is of the order of $10^{13}$ atoms/cm$^{3}$. The PA experiment is performed at a temperature of around 4 $\mu$K, just above the critical temperature of Bose-Einstein condensation~\cite{Robert,He-BEC1} to avoid mean field shifts.

The PA light is provided by a DBR diode laser with an external cavity operating at 1083 nm and a 1 W Yb-doped fiber amplifier. The PA pulse is focused with a waist of 640 $\mu$m on the atomic sample of approximately 30 $\mu$m diameter (in radial direction) and switched on for durations from tens of $\mu$s to a few ms depending on the laser intensity. The light propagates parallel to the trap bias magnetic field in $\sigma^-$ polarization to produce selectively the desired bound states from v=0 to v=4 in the purely long-range $0_u^+,~J=1$ potential by setting its detuning in the range of -20 MHz to -1.4 GHz~\cite{PA-He,Kim}. In order to allow large detunings, we use a heterodyne frequency locking technique: the beat signal between the PA laser and a master laser is monitored by a fast photodiode, mixed down with a local oscillator and converted to DC voltage which is then fed back to the current supply of the PA laser. The master laser is locked to the atomic D$_0$ transition. The linewidth of the PA laser is measured to 0.6 MHz. Our polarization purity is limited to approx. 98 \% in the electric field because of imperfections in the optical elements.

\begin{figure}
\begin{center}
\centerline{\includegraphics[scale=0.30]{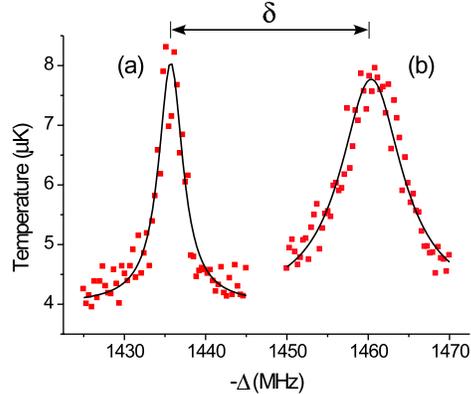}}
\end{center}
\caption{Two PA resonance curves of the lowest ($v=0$) vibrational level of the $0_u^+$ potential with PA laser intensity of (a) 9 mW/cm$^2$ and (b) 5 W/cm$^2$ and pulse duration of 1 ms and 10 $\mu$s respectively. The heating of the cloud due to the PA process is plotted as a function of the opposite of the laser detuning $\Delta$ from the $2^3S_1-2^3P_0$ transition. The two sets of data are fitted with lorentzian curves represented in solid lines, and the center frequencies are found to be (a) -1435.0 $\pm$ 0.6 MHz and (b) -1460 $\pm$ 1 MHz. The resonance peak (b) is clearly shifted by a quantity $\delta$ larger than 20 MHz at strong PA laser intensity.}
\label{fig:shift-raw}
\end{figure}

The PA resonance curves are obtained by the calorimetric method described in Ref.~\cite{PA-He}. Even for intensities several order of magnitudes higher in the present work, the rise in temperature, the optical density and the atom loss signals (which are simultaneously registered) still give the same central position of the resonance line. This validates our detection method. We deduce the temperature by ballistic expansion. For each PA line, the PA laser frequency is scanned over 20 MHz in steps of 0.3 MHz by changing the RF-frequency of an acousto-optic modulator in the PA beam line. 

Fig.~\ref{fig:shift-raw}(a) shows a PA resonance line obtained at low PA laser intensity. The width is typically 3 MHz. When the PA laser intensity is increased, the line is red shifted and broadened as shown in Fig.~\ref{fig:shift-raw}(b). We decrease the pulse duration as we raise the laser intensity in order to keep enough atoms trapped after photoassociation, so that the heating by the PA process is still measurable. We checked that the resonance position is  independent of the pulse duration by varying it by more than an order of magnitude. The amplitude of the temperature raise signal is not proportional to the laser energy: when the laser energy focused on the cloud is large, the atom loss is not negligible compared to the initial number of atoms, and the thermalization properties of the cloud are changed. With the temperature and laser intensity we use, the broadening of the signal is too large to be explained by Ref. \cite{Bohn99,Simoni} where it is related to the transition rate from the coupled scattering states to the excited state. However, some decay mechanisms enhanced by light are suggested by the observation of ions when a metastable helium cloud is illuminated by a laser red-detuned from the atomic transition \cite{Woestenenk,Rijnbach}.

\section{Frequency shifts measurements}

\begin{figure}
\onefigure[scale=0.40]{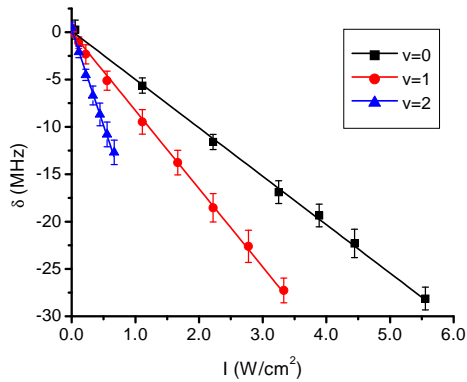}
\caption{Measurements of the light-induced frequency-shift $\delta$ of the PA resonance as a function of the PA laser intensity I for different molecular bound states (v=0,1,2) of the $0_u^+$ potential. The ratio between two slopes achieved in same experimental conditions ($\delta_{\mathrm{v}=1} / \delta_{\mathrm{v}=0}$ or $\delta_{\mathrm{v}=2} / \delta_{\mathrm{v}=0}$ ) is used to compare with theory to provide a value for the scattering length $a$.}
\label{fig:shift-comp}
\end{figure}

For each vibrational level, we measure the shifts $\delta$ of the
center of the resonance curve for different PA laser intensities
I. Typical results are shown in Fig.~\ref{fig:shift-comp}. The
error bar is explained as follows. At high intensities, the error
bar in $\delta$ increases due to signal broadening and large
shot-to-shot noise. Close to zero intensity, we assign an error of
0.6 Mhz to $\delta$ resulting from fluctuations of experimental
parameters, which is larger than the error of the Lorentzian fit.

The results displayed in Fig.~\ref{fig:shift-comp} show a linear dependance of $\delta_{\mathrm{v}=0,1,2}$ as a function of I. The maximum measurable light shift is limited by two experimental factors. First, the signal broadens with the laser intensity. Second, to keep the amount of loss of atoms low enough, the pulse duration is decreased when I is increased as mentioned earlier, which is limited by the 5 $\mu$s switching time of the AOM. These constraints are more severe for v=2.
The main error in the determination of the slopes $\delta_{\mathrm{v}}$ is due to the systematic uncertainty in the intensity I.  This results from the difficulty to measure precisely the beam diameter at the cloud location, the calibration of our laser power meter, a possible slight misalignment of the laser on the cloud and the optical losses on the glass cell walls. We estimate that these factors lead to an error of up to 50~\% in the absolute value of I. This is why our measurement of the scattering length relies on the ratio of the slopes for two vibrational levels : v=0 and v=1 or v=0 and v=2, which is not affected by the systematic error in the intensity.
We also observed that the laser intensity to which we expose the atomic cloud varies from day to day by approximately 10~\%, due to small changes of the position of the center of the magnetic trap. The slopes of the three levels are therefore measured in a single run without interruption to keep the experimental conditions as similar as possible.

The precision in the measurement of each ratio is limited by the fitting errors in the slopes $\delta_\mathrm{v}$ which are 2~\% for v=0, 2.5~\% for v=1, and 4.5~\% for v=2 (see Fig.~\ref{fig:shift-comp}). The ratios of the frequency shifts are measured several times, and we find the results $1.71 \pm 0.14$ for $\delta_{\mathrm{v}=1} / \delta_{\mathrm{v}=0}$ and $4.20 \pm 0.48$ for $\delta_{\mathrm{v}=2} / \delta_{\mathrm{v}=0}$. The error bars are chosen such that they include all the measurements and their individual error bars, and a statistical  analysis shows that they correspond to a three standard deviation error. The result for $\delta_{\mathrm{v}=1} / \delta_{\mathrm{v}=0}$ is the most precise because larger shifts produce less error in the linear fit.

\section{Theory}
The details of the theoretical analysis which provides the relation between the shifts and the scattering length for different vibrational levels and light polarizations will be given in great detail in a forthcoming paper \cite{Portier}. We focus here on its main features. 

We consider the light coupling of the molecular bound state in the purely long-range $J=1;M_J=1;0_u^+$ potential to the continuum of scattering states in the $l=0, 2, 4$ waves as well as in several spin states allowed by selection rules. Under our experimental conditions (ultracold atoms in a magnetic trap), the only significant entrance collisional channel corresponds to two spin-polarized atoms colliding in the s-wave. However, the contribution to the light shift of all other possible allowed channels may not be negligible \cite{PA-Na-BEC}. The fact that $l\neq0$ channels contribute to the shift suggests a dependance on light polarization. In addition, the contributions of the ground molecular bound states is included. The least bound state v=14 of the $^5\Sigma_g^+$ potential gives the most important one, as the other vibrational levels are much further from resonance. 

Using Ref. \cite{Simoni}, it can be shown that the shift $\delta_\mathrm{v}$ is the sum of partial shifts $\delta_\mathrm{v}^{(i)}$ due to different collisionnal channels $(i)$. Only the partial shifts corresponding to s-wave collisionnal channels depend on the scattering length. Each $\delta_\mathrm{v}^{(i)}$ can be written as a sum over bound and continuum states or as an overlap integral involving the regular and irregular solutions of the scattering equation, and the excited state wavefunction \cite{Bohn99}. The former can be used to show that in our case  both contributions of the least-bound state and the scattering states in the s-wave collisional channel provide the sensitivity of the shift on $a$. The latter is preferred for computational reasons, and proves that the shift is a linear function of $a$, very weakly dependent on the scattering energy $E_{sc}$ and on the magnetic field for the low temperature and values of the field we use. We prefer using experimentally the ratio of the shifts between different vibrational levels which is independent on intensity. In our case, the fraction of the shift which depends on $a$ is appreciably different for the different levels. The ratios of the shifts depend therefore in a critical way on $a$.

Several approximations are made in the theoretical analysis. First, the isolated level approximation, which takes into account that the resonances associated to the excited molecular states are well separated, is valid as their shifts and broadenings are much smaller than the level spacing. Second, the different ground state collisional channels can be assumed to be  uncoupled in the absence of light, as the spin-dipole interaction, which mixes different spin states, is very weak \cite{Fedichev}. Third, we neglect dressing effects, which arise from the fact that at infinite separation the colliding atoms still interact with light \cite{Napolitano}. As we solely deal with the deepest bound excited states our laser is tuned far from the atomic resonance. Although these approximations are made, the errors they induce on the determination of $a$ are much smaller than the one due to experimental uncertainty on the ratio measurement.

\section{Scattering length}

\begin{figure}
\begin{center}
\resizebox{0.9\columnwidth}{!}{
\includegraphics[scale=1]{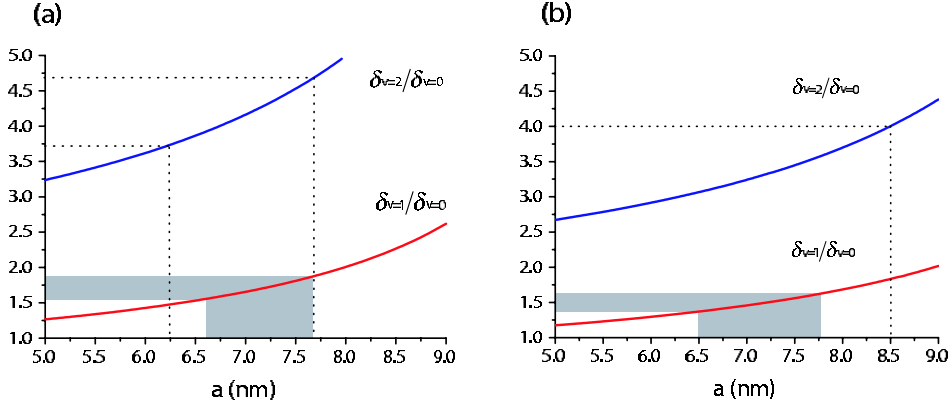}}
\end{center}
\caption{Ratios of the frequency-shift for the PA lines obtained
with $2^3S_1$ helium atoms excited to v=0, 1, and 2 in the $0^+_u$
potential. Fig. (a) shows the result with $\sigma^-$ polarization
: $\delta_{\mathrm{v}=1} / \delta_{\mathrm{v}=0} = 1.70 \pm 0.14$, $\delta_{\mathrm{v}=2} /
\delta_{\mathrm{v}=0} = 4.2 \pm 0.5$. and Fig. (b) with linear polarization: $\delta_{\mathrm{v}=1} /
\delta_{\mathrm{v}=0} = 1.50 \pm 0.12$, $\delta_{\mathrm{v}=2} / \delta_{\mathrm{v}=0} = 3.3
\pm 0.7$. The curves for $\delta_{\mathrm{v}=1} / \delta_{\mathrm{v}=0}$ and
$\delta_{\mathrm{v}=2} / \delta_{\mathrm{v}=0}$ are provided by our theoretical
calculation~\cite{Portier}. The experimental uncertainty is represented by the grey area for the $\delta_{\mathrm{v}=1} / \delta_{\mathrm{v}=0}$ ratio, and the dashed lines for the $\delta_{\mathrm{v}=2} / \delta_{\mathrm{v}=0}$ ratio. In Fig. (b) the lower
end of the $\delta_{\mathrm{v}=2} / \delta_{\mathrm{v}=0}$ is 2.6, out of the plotted range for $a$. The value for the scattering length of $a=7.2\pm0.6$~nm is provided by the measurement for the $\delta_{v=1}/\delta_{v=0}$ ratio with $\sigma^-$ polarization.}
\label{fig:ratios}
\end{figure}

We compare the measured ratio of the light shift for $\delta_{\mathrm{v}=1} / \delta_{\mathrm{v}=0}$ with the calculations in Ref.~\cite{Portier}. The fact that the polarization is not pure is taken into account. As mentioned above, our theory does not add uncertainty when deducing the scattering length. The measured ratio of the light shift for $\delta_{\mathrm{v}=1} / \delta_{\mathrm{v}=0}$ provides a determination of the scattering length of $a=7.2 \pm 0.5$~nm as shown in figure~\ref{fig:ratios}(a). The ratio $\delta_{\mathrm{v}=2} / \delta_{\mathrm{v}=0}$ and the measurements for linear polarization shown in figure~\ref{fig:ratios}(b) provide less precise but yet compatible values. One can infer from the present determination of $a$ the absolute values of the slopes of $\mathrm{\delta_v}$ as a function of I for different polarizations (see Ref.~\cite{Portier}). We find a systematic discrepancy of 40\% with the experimental results, which are smaller in absolute value. We attribute this to an overestimation of the intensity I.

This new value of $a$ can be compared with the previous experimental determinations. In most cases the error-bars do not overlap. The condensate expansion yields  $a=20 \pm 10$~nm~\cite{Robert} and $a=16 \pm 8$~nm~\cite{He-BEC1}. The largest discrepancy is found with the value of the Orsay group $a=11.3 \: ^{+2.5}_{-1.0}$~nm derived from inelastic collision rates combined with the determination of the critical temperature~\cite{Seidelin}. However, the measurement of the elastic collision rate based on the evaporative cooling rate gave $a=10\pm5$~nm~\cite{Tol}, which is in good agreement with our result.  The small value of $a$ could explain why the hydrodynamic regime was not reached in our previous experiment accounting for the collective excitation of collisionally dense thermal clouds~\cite{Leduc}. The comparison with theoretical values obtained from ab~initio calculations is also instructive. Using the potential of Starck and Meyer~\cite{Starckmeyer} a value of $a = 8.3$ nm is found. More recently, a value of 8.0 nm $<a<$ 12.2 nm has been calculated~\cite{Gadea2004}.  After completion of the present work, we became aware of the result of Przybytek and Jeziorski~\cite{Polish}, which agrees very well with our measurement.

\section{Conclusion}

We measure precisely the light-induced frequency shift of the PA spectra of metastable Helium atoms and compared our results with our theoretical calculation. The measurements of several ro-vibrational levels at different light polarizations are considered. We also provide the first PA spectroscopic measurement of the s-wave scattering length of spin-polarized Helium atoms in metastable $2^3S_1$ state. The value for $a$ presented in this letter is significantly lower as compared to most previous values.

A more precise spectroscopic method to determine the scattering length would be to measure the binding energy of the least bound state (v=14) in the ground state $^5\Sigma_g^+$ potential with a two-photon scheme. This PA experiment is in progress in our group.

\acknowledgments

The authors thank Olivier Dulieu from the Cold Atoms and Molecules
group at Laboratoire Aime Cotton in France and Paul Julienne from
National Institute of Standards and Technology in USA, for
fruitful discussions.

\end{document}